\input harvmac

%
\let\includefigures=\iftrue
%
%
%
\newfam\black
\input rotate
\input epsf
\input psfig
\noblackbox
%
%
\includefigures
\message{If you do not have epsf.tex (to include figures),}
\message{change the option at the top of the tex file.}
\def\figin{\epsfcheck\figin}\def\figins{\epsfcheck\figins}
\def\epsfcheck{\ifx\epsfbox\UnDeFiNeD
\message{(NO epsf.tex, FIGURES WILL BE IGNORED)}
\gdef\figin##1{\vskip2in}\gdef\figins##1{\hskip.5in}
\else\message{(FIGURES WILL BE INCLUDED)}%
\gdef\figin##1{##1}\gdef\figins##1{##1}\fi}
\def\DefWarn#1{}

\def\figinsert{\goodbreak\midinsert}
\def\ifig#1#2#3{\DefWarn#1\xdef#1{fig.~\the\figno}
\writedef{#1\leftbracket fig.\noexpand~\the\figno}%
\figinsert\figin{\centerline{#3}}\medskip\centerline{\vbox{\baselineskip12pt
\advance\hsize by -1truein\noindent\footnotefont{\bf
Fig.~\the\figno:} #2}}
\bigskip\endinsert\global\advance\figno by1}
\else
\def\ifig#1#2#3{\xdef#1{fig.~\the\figno}
\writedef{#1\leftbracket fig.\noexpand~\the\figno}%
\global\advance\figno by1} \fi
\def\yboxit#1#2{\vbox{\hrule height #1 \hbox{\vrule width #1
\vbox{#2}\vrule width #1 }\hrule height #1 }}
\def\fillbox#1{\hbox to #1{\vbox to #1{\vfil}\hfil}}
\def\ybox{{\lower 1.3pt \yboxit{0.4pt}{\fillbox{8pt}}\hskip-0.2pt}}

\def\rightarrowbox#1#2{
  \setbox1=\hbox{\kern#1{${ #2}$}\kern#1}
  \,\vbox{\offinterlineskip\hbox to\wd1{\hfil\copy1\hfil}
    \kern 3pt\hbox to\wd1{\rightarrowfill}}}

\def\half{{1\over 2}}
\def\Tr{{{\rm Tr~ }}}

\def\vev#1{\langle{#1}\rangle}

\def\tilde{\widetilde}

\def\II{\relax{I\kern-.10em I}}

\def\pt{pt}
\def\bar{\overline}

\def\IZ{\relax\ifmmode\mathchoice
{\hbox{\cmss Z\kern-.4em Z}}{\hbox{\cmss Z\kern-.4em Z}}
{\lower.9pt\hbox{\cmsss Z\kern-.4em Z}} {\lower1.2pt\hbox{\cmsss
Z\kern-.4em Z}}\else{\cmss Z\kern-.4em Z}\fi}
\def\IB{\relax{\rm I\kern-.18em B}}
\def\IC{{\relax\hbox{$\inbar\kern-.3em{\rm C}$}}}
\def\ID{\relax{\rm I\kern-.18em D}}
\def\IE{\relax{\rm I\kern-.18em E}}
\def\IF{\relax{\rm I\kern-.18em F}}
\def\IG{\relax\hbox{$\inbar\kern-.3em{\rm G}$}}
\def\IGa{\relax\hbox{${\rm I}\kern-.18em\Gamma$}}
\def\IH{\relax{\rm I\kern-.18em H}}
\def\II{\relax{\rm I\kern-.18em I}}
\def\IK{\relax{\rm I\kern-.18em K}}
\def\IN{\relax{\rm I\kern-.18em N}}
\def\IP{\relax{\rm I\kern-.18em P}}

%
\def\inbar{\,\vrule height1.5ex width.4pt depth0pt}

\font\cmss=cmss10 \font\cmsss=cmss10 at 7pt
\def\IR{\relax{\rm I\kern-.18em R}}

\def\lp10{l_P^{10}}
\def\lp11{l_P^{11}}
\def\R11{R_{11}}

\newbox\tmpbox\setbox\tmpbox\hbox{\abstractfont
}
 \Title{\vbox{\baselineskip12pt\hbox to\wd\tmpbox{\hss
 hep-th/0403047} }}
 {\vbox{\centerline{MHV Vertices}
 \smallskip
 \centerline{and Tree Amplitudes In Gauge Theory}
 }}
\smallskip
\centerline{Freddy Cachazo,$^a$ Peter Svrcek,$^b$ and Edward Witten$^a$}
\smallskip
\bigskip
\centerline{\it $^a$ School of Natural Sciences, Institute for Advanced
Study, Princeton NJ 08540 USA}
\bigskip
\centerline{\it $^b$ Joseph Henry Laboratories, Princeton
University, Princeton NJ 08544 USA}
\bigskip
\vskip 1cm \noindent

\input amssym.tex
As an alternative to the usual Feynman graphs, tree amplitudes in
Yang-Mills theory can be constructed from tree graphs in which the
vertices are tree level MHV scattering amplitudes, continued off
shell in a particular fashion.  The formalism leads to new and
relatively simple formulas for many amplitudes, and can be
heuristically derived from twistor space.

\Date{March 2004}
%
\newsec{Introduction}
\nref\dewitt{B. DeWitt, ``Quantum Theory Of Gravity, III: Applications Of
The Covariant Theory,'' Phys. Rev. {\bf 162} (1967) 1239.}%
\nref\pt{S. Parke and T. Taylor, ``An Amplitude For $N$ Gluon Scattering,''
Phys. Rev. Lett. {\bf 56} (1986) 2459.}%
\nref\bg{F. A. Behrends and W. T. Giele, ``Recursive Calculations For Processes
With $N$ Gluons,'' Nucl. Phys. {\bf B306} (1988) 759.}%
\nref\bkd{Z. Bern, L. Dixon, and D. Kosower, ``Progress In One-Loop QCD
Calculations,'' hep-ph/9602280, Ann. Rev. Nucl. Part. Sci. {\bf 36} (1996) 109.}%
\nref\newbern{C. Anastasiou, Z. Bern, L. Dixon, and D. Kosower, ``Planar
Amplitudes In Maximally Supersymmetric Yang-Mills Theory,'' hep-th/0309040;
Z. Bern, A. De Freitas, and L. Dixon, ``Two Loop Helicity Amplitudes For Quark
Gluon Scattering In QCD and Gluino Gluon Scattering In Supersymmetric Yang-Mills
Theory,'' JHEP 0306:028 (2003), hep-ph/0304168.}%

Perturbative scattering amplitudes in Yang-Mills theory have
remarkable properties that are not apparent from the textbook
recipes for computing them.  Unexpected selection rules for
helicity amplitudes were uncovered in the earliest computation of
tree level gluon scattering \dewitt. Tree amplitudes in which the
maximal number of gluons have the same helicity are described by a
marvelously simple formula \refs{\pt,\bg}. (These are known as
maximal helicity violating or MHV amplitudes.) Loop amplitudes
also turn out to be unexpectedly simple \refs{\bkd,\newbern}.

Some properties of perturbative Yang-Mills theory may apparently
be explained \ref\wittwistors{E. Witten, ``Perturbative Gauge
Theory As A String Theory In Twistor Space,'' hep-th/0312171.} by
relating this theory to the instanton expansion of a certain
string theory in twistor space \ref\penrose{R. Penrose, ``Twistor
Algebra,'' J. Math. Phys. {\bf 8} (1967) 345.}. In the present
paper, we reconsider the tree amplitudes of perturbative
Yang-Mills theory in a way that is suggested by the twistor
transform (and by our study of differential equations obeyed by
scattering amplitudes, which will appear elsewhere) and also by
the use \refs{\bkd,\newbern} of MHV tree amplitudes in calculating
loop amplitudes.

Consider a theory in Minkowski space of gauge invariant local
fields such as scalar fields $\phi_i$.  We consider a local
interaction vertex such as a polynomial interaction $W = \int
d^4x\, F(\phi_i)$. A point in Minkowski space corresponds
\penrose\ to a ``line'' in twistor space -- that is, to a linearly
embedded copy of $\Bbb{CP}^1$.  So the interaction vertex
$F(\phi_i)$, which is supported on a point in Minkowski space, is
supported on a line in twistor space.

As shown in \wittwistors, the tree level MHV amplitudes for
scattering of any number of gluons of positive helicity and two of
negative helicity is similarly supported on a line.  So we think
of this amplitude as representing, in some sense, a generalization
of a local interaction vertex.  This is in the spirit of analyses
of loop diagrams \refs{\bkd,\newbern} in which, roughly speaking,
MHV tree level amplitudes are regarded as interactions and
amplitudes that are rational functions (of the spinor variables
used to describe external particles) are considered to be
``local.''

In this paper, we pick a specific off-shell continuation of the
MHV amplitude and consider Feynman diagrams in which the vertices
are tree level MHV amplitudes -- with an arbitrary number of gluon
lines -- and the propagator is the standard Feynman propagator
$1/p^2$.  We call these diagrams MHV diagrams.

Our off-shell continuation of the MHV amplitude is not
Lorentz-covariant, and the sum of  MHV diagrams is not manifestly
Lorentz-covariant.  Nevertheless, we argue that the sum of MHV
tree diagrams is covariant, and we verify, for examples with five,
six, or seven gluons, that this sum coincides with conventional
Yang-Mills tree amplitudes.

Assuming that this is so to all orders, we obtain relatively short
and simple expressions for certain amplitudes, such as the
helicity amplitudes $---+++\dots+$. See \ref\kosower{D. Kosower,
``Light Cone Recurrence Relations For QCD Amplitudes,'' Nucl.
Phys. {\bf B335} (1990) 23.} for previously known formulas for
these amplitudes.

In section 2, we describe our off-shell continuation.  In section
3, we describe explicit computations of some amplitudes. In
section 4, we verify that MHV tree amplitudes have the same
collinear and multiparticle singularities as the standard
Yang-Mills tree amplitudes.  In section 5, we prove that the sum
of MHV tree amplitudes is Lorentz-covariant.  Finally, in section
6, we attempt to justify the MHV tree amplitudes as a method of
evaluating the twistor amplitudes coming from completely
disconnected instantons (that is, from collections of disjoint
instantons each of which has instanton number one). This argument
is not really rigorous as the rules for what integration contours
to use in twistor space are not entirely clear.

The argument in section 6  raises a puzzle to which we do not have
an answer.   Other recent results suggest that it is possible to
compute the same amplitudes solely from connected instantons
\ref\rsv{R. Roiban, M. Spradlin, and A. Volovich, ``A Googly
Amplitude From The $B$ Model In Twistor Space,'' hep-th/0402016;
R. Roiban and A. Volovich, ``All Googly Amplitudes From The $B$
Model In Twistor Space,'' hep-th/0402121.}.   Why  might it be
possible to compute the same amplitudes from connected instantons
or from completely disconnected ones?  Perhaps in some topological
string theory, it is possible to choose one integration contour in
field space that picks up only the connected instantons and
another one that picks up only the completely disconnected
instantons.

\newsec{Definition Of MHV Tree Amplitudes}

We recall that in four dimensions, a momentum vector $p_\mu$ can
conveniently be represented as a bispinor $p_{a\dot a}$ and that
the momentum vector for a massless particle can be factored as
$p_{a\dot a}=\lambda_a\tilde\lambda_{\dot a}$ in terms of spinors
$\lambda_a$, $\tilde\lambda_{\dot a}$ of positive and negative
chirality. Spinor inner products are denoted as $\langle
\lambda,\lambda'\rangle=\epsilon_{ab}\lambda^a\lambda'{}^b$,
$[\tilde\lambda,\tilde\lambda']=\epsilon_{\dot a\dot
b}\tilde\lambda^{\dot a}\tilde\lambda'{}^{\dot b}$.  If $p_{a\dot
a}=\lambda_a \tilde\lambda_{\dot a}$, $q_{a\dot
a}=\lambda'_a\tilde\lambda'_{\dot a}$, then $2p\cdot
q=\langle\lambda,\lambda'\rangle[\tilde\lambda,\tilde\lambda']$.
For more detail and references, see \wittwistors.

We will be studying tree level scattering amplitudes with $n$
gluons.  Such an amplitude is in a natural way a sum of
subamplitudes associated with different cyclic orderings of the
external gluons; we focus on the term associated with a particular
cyclic ordering, say the one for which the group theory factor is
$\Tr\,T_1T_2\dots T_n$.  We suppress this factor in writing the
amplitudes.

A tree level scattering amplitude with $n$ gluons more than $n-2$
of which have the same helicity vanishes.  The amplitudes with
$n-2$ gluons all of the same helicity are called maximally
helicity violating or MHV amplitudes.  The MHV tree amplitude with
$n-2$ gluons of positive helicity are as follows \refs{\pt,\bg}.
If the gluons of negative helicity are labeled $x,y$ (which may be
any integers from $1$ to $n$), the amplitude is
\eqn\tonno{A_n={\langle\lambda_x,\lambda_y\rangle^4\over
\prod_{i=1}^n\langle\lambda_i, \lambda_{i+1}\rangle }.} (We omit
the trace $\Tr\,T_1\dots T_n$, a delta function
$(2\pi)^4\delta^4(\sum_i\lambda_i^a\tilde\lambda_i^{\dot a})$ of
energy-momentum conservation, and a factor $g^{n-2}$, with $g$ the
Yang-Mills coupling.)

In this paper, we will continue these ``mostly plus'' MHV
amplitudes off-shell and use them as vertices in tree diagrams
that we will call MHV tree diagrams.\foot{A ``mostly plus'' MHV
amplitude has two gluons of negative helicity and any number of
positive helicity.  In the exceptional cases that the number of
positive helicity gluons is one or two, there is not really a
majority of gluons with positive helicity.} (We do not include
additional vertices for the ``mostly minus'' MHV tree amplitudes;
along with other amplitudes, they are computed from trees with
``mostly plus'' vertices.) In the physical amplitude \tonno, each
particle is assumed to be on-shell, with lightlike momentum vector
$p_{\,a\dot a}=\lambda_{\,a}\tilde\lambda_{\dot a}$.  To
generalize the MHV tree amplitude to a vertex that can be inserted
in a Feynman diagram, we need to continue it off-shell. An
off-shell field is still characterized by a momentum vector
$p_{a\dot a}$, but what can be meant by $\lambda_a$ if $p$ is not
lightlike?

Suppose that $p_{a\dot a}$ is lightlike.  We can pick an arbitrary
negative chirality spinor $\eta^{\dot a}$ and then up to scaling
we can take $\lambda_a=p_{a\dot a}\eta^{\dot a}$.  In fact, if
$p_{a\dot a}=\lambda_a\tilde\lambda_{\dot a}$, then $\lambda_a
=p_{a\dot a}\eta^{\dot a}/[\tilde\lambda,\eta]$.  The factor
$1/[\tilde\lambda,\eta]$ is irrelevant since tree amplitudes that
we compute will always be invariant under rescaling of the
$\lambda$'s for all the off-shell, internal lines.

This leads to our definition of the off-shell continuation.  We
simply pick an arbitrary $\eta^{\dot a}$ and then define
$\lambda_a$ for any internal line carrying momentum $p_{a\dot a}$
in a Feynman diagram by \eqn\tonron{\lambda_a=p_{a\dot
a}\eta^{\dot a}.} For example, if $\eta^{\dot a}=\delta^{\dot a
2}$, then the definition is  $\lambda_a=p_{a\dot 2}$. We use the
same $\eta$ for all the off-shell lines in all diagrams
contributing to a given amplitude. For external lines -- lines
representing incoming or outgoing gluons in a scattering process
-- $\lambda$ is defined in the usual way in terms of the wave
function of the initial or final particle. With this understanding
of what $\lambda$ means for each particle, we simply take the
``mostly plus'' MHV scattering amplitude \tonno\ as the $n$-gluon
vertex in our Feynman diagram, for all $n\geq 3$. (We introduce no
additional vertices for the ``mostly minus'' MHV amplitudes. They
will be computed from tree diagrams using mostly plus vertices, as
we see in the next section.)

At each interaction vertex, each gluon, understood to be incoming,
is assigned a definite helicity.  This is so for both on-shell and
off-shell lines.  In fact,  at an $n$-gluon  vertex, $n-2$ of the
gluons have positive helicity and two have negative helicity; in
\tonno, the two gluons of negative helicity have been labeled
$x,y$. If a gluon is considered to be outgoing, its helicity label
is reversed.

For the propagator of an off-shell gluon of momentum $p$, we take
simply $1/p^2$.  The two ends of any propagator must have opposite
helicity labels -- plus at one end and minus at the other end --
because an incoming gluon of one helicity is equivalent to an
outgoing gluon of the opposite helicity.

\ifig\tramp{A tree diagram with MHV vertices.  In this example, the number of
vertices is $v=3$; they are connected by $v-1=2$ propagators.
The vertices are respectively trivalent, four-valent, and
five-valent.  Internal and external lines are labeled by their
helicity.} {\epsfbox{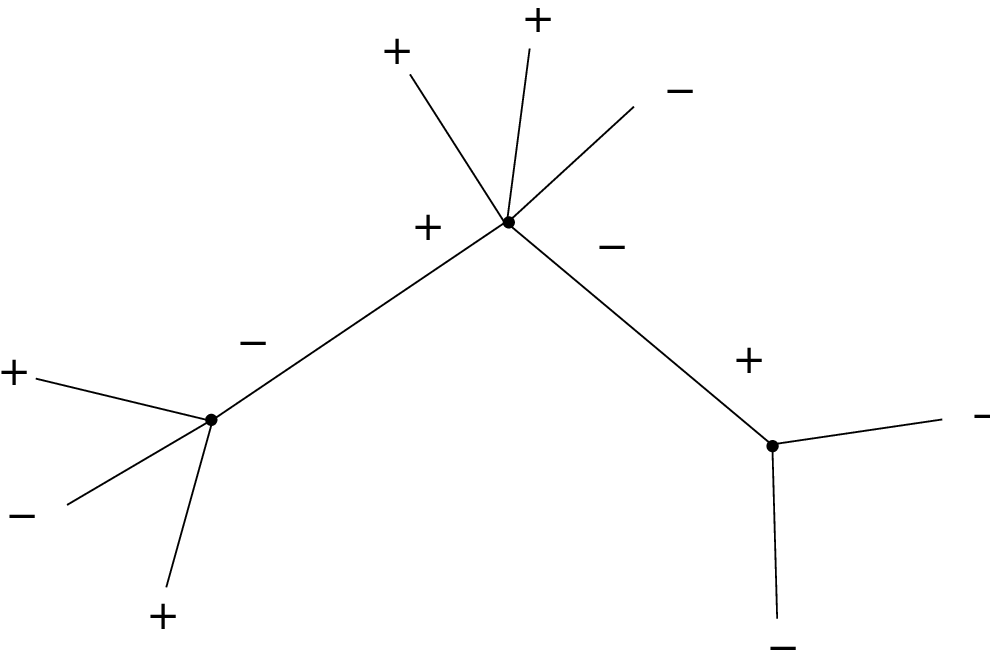}}

 In the next section, we give some examples of Feynman diagrams
computed using these rules.  For now, we simply make the following
observation.  Consider a tree diagram with $v$ vertices and
(therefore) $v-1$ propagators connecting them.  (See figure 1 for
an example.)  Each vertex has precisely two gluon lines of
negative helicity emanating from it. So a total of $2v$ negative
helicity gluon lines emanate from the vertices.  Each of the $v-1$
propagators connects at precisely one end to one of these $2v$
lines. This leaves $v+1$ negative helicity lines that must be
attached to external particles.  In other words,  a tree level
scattering amplitude with  $q$ external gluons of negative
helicity must be obtained from an MHV tree diagram with $v$
vertices such that $q=v+1$ or equivalently \eqn\moggo{v=q-1.} This
implies, in particular, that MHV tree diagrams with $q<2$ external
gluons vanish, since they contain no vertices at all.  This is in
agreement with the fact that these amplitudes vanish in Yang-Mills
theory.  Moreover, if $q=2$, the number of vertices is $v=1$, and
the MHV tree amplitude is equal by definition to the Yang-Mills
tree amplitude.  The first nontrivial case of our claim is for
$q=3$, $v=2$.

The result \moggo\ is analogous to the result in \wittwistors\
that a Yang-Mills tree amplitude with $q$ gluons of negative
helicity (and any number of positive helicity) must be derived
from a curve in twistor space of degree or instanton number
$d=q-1$.  The degree one curves in twistor space correspond to the
MHV vertices in the present approach.

\newsec{Examples}

Here we will describe examples of evaluation of MHV tree
amplitudes. As just explained, the first case to consider is that
the number of negative helicity gluons is $q=3$ and the number of
vertices is therefore $v=2$.

\ifig\poramp{MHV tree diagrams contributing to the $+---$
amplitude, which is expected to vanish.  Arrows indicate the
momentum flow, while $+$ and $-$ signs denote the helicity. }
{\epsfbox{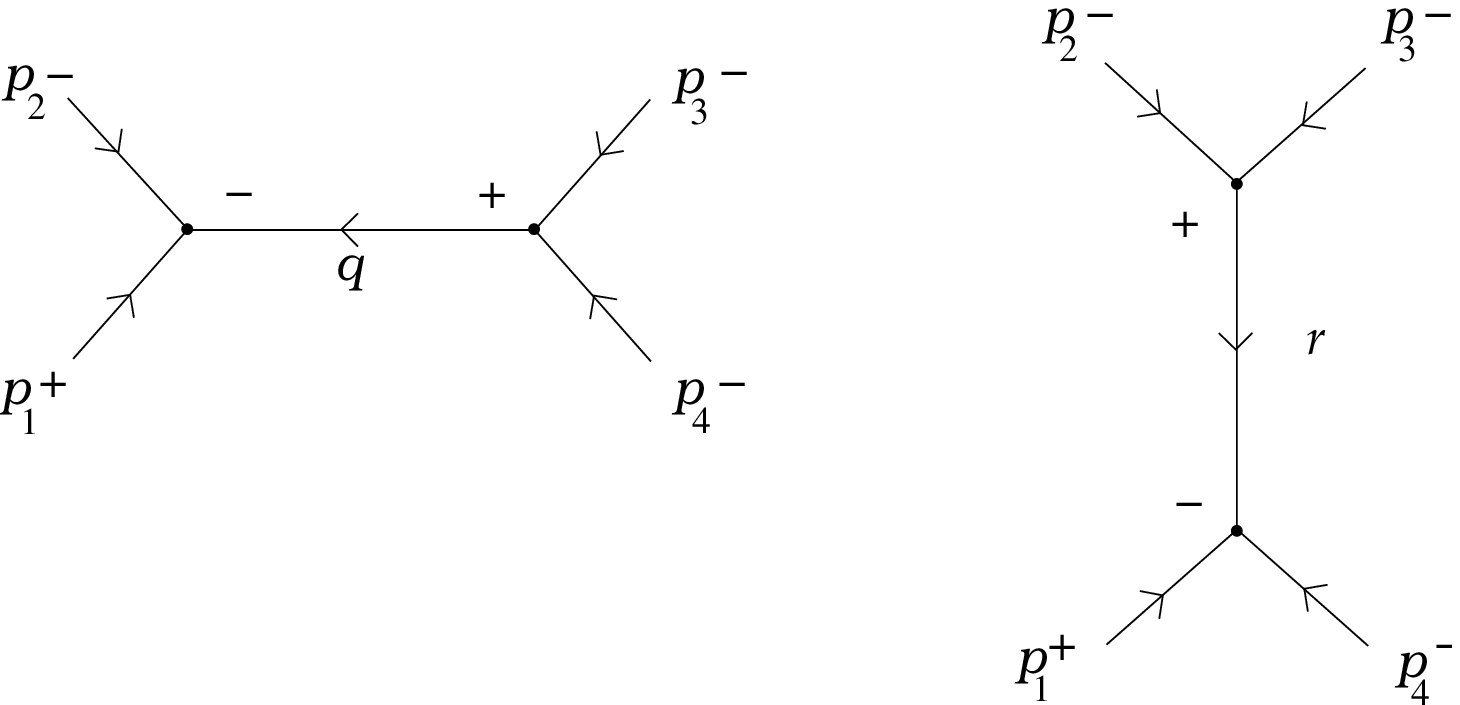}}

We begin with the first case, the four gluon amplitude with
momenta $p_1,\dots,p_4$ and helicities $+--\,-$.  This vanishes in
Yang-Mills theory; we want to verify that it also vanishes when
computed from MHV tree diagrams.  As indicated in figure 2, there
are two diagrams to consider. In the first diagram, there is an
internal line with momentum $q=-p_1-p_2=p_3+p_4$. We write
$\lambda_q,$ $\tilde\lambda_q$ for the corresponding spinors. As
explained in section 2, $\lambda_{q\,a}=q_{a\dot a}\eta^{\dot a}$,
for some arbitrary $\eta^{\dot a}$ (which we take to be the same
in both diagrams). We abbreviate $\tilde\lambda_{i\,\dot
a}\eta^{\dot a}$ as $\phi_i$. So
\eqn\onsonn{\lambda_{q\,a}=-\lambda_{1\,a}\phi_1
-\lambda_{2\,a}\phi_2=\lambda_{3\,a}\phi_3 +\lambda_{4\,a}\phi_4,}
where we have used the fact that $p_{i\,a\dot a}=\lambda_{i\,a}
\tilde\lambda_{i\,\dot a}$.
The amplitude associated with the first diagram in figure 2 is
\eqn\onop{{\langle\lambda_2,\lambda_q\rangle^3\over
\langle\lambda_q,\lambda_1\rangle\langle\lambda_1,\lambda_2\rangle}
{1\over q^2}{\langle\lambda_3,\lambda_4\rangle^3\over
\langle\lambda_4,\lambda_q
\rangle\langle\lambda_q,\lambda_3\rangle}.} To obtain this
formula, we took the propagator to be $1/q^2$, and we read off the
trivalent vertices from \tonno.

{}From \onsonn, we have $\langle\lambda_2,
\lambda_q\rangle=-\langle 2~1\rangle \phi_1$,
$\langle\lambda_q,\lambda_1\rangle=-\langle 2~1\rangle \phi_2$,
$\langle \lambda_4,\lambda_q\rangle=\langle 4~3\rangle\phi_3$, and
$\langle\lambda_q,\lambda_3\rangle=\langle 4~3\rangle\phi_4$. (We
recall that $\langle i~j\rangle$ is an abbreviation for
$\langle\lambda_i,\lambda_j\rangle$.) So \onop\ becomes
\eqn\tonop{{\phi_1^3\over\phi_2\phi_3\phi_4}{\langle
2~1\rangle^3\over \langle 2~1\rangle \langle 1 ~2\rangle}{1\over
q^2}{\langle 3~4\rangle^3\over \langle 4~3\rangle
\langle4~3\rangle}.} Using $q^2=(p_1+p_2)^2=2p_1\cdot p_2=\langle
1~2\rangle[1~2]$, and $\langle i~j\rangle=-\langle j~i\rangle$,
this becomes \eqn\gonop{-{\phi_1^3\over\phi_2\phi_3\phi_4}{\langle
3~4\rangle\over[2~1]}.} A very similar evaluation of the second
diagram gives \eqn\bonop{-{\phi_1^3\over
\phi_2\phi_3\phi_4}{\langle 3~2\rangle\over [4~1]}.} In fact, a
new evaluation is not needed, since the second diagram can be
obtained from the first by exchanging particles 2 and 4.  The sum
of these expressions vanishes, since momentum conservation implies
that $0=\sum_i \langle 3~i\rangle[i~1]=\langle
3~2\rangle[2~1]+\langle 3~4\rangle [4~1].$

\ifig\truramp{(a) MHV tree diagrams contributing to the $+-+--$
amplitude.  (b) This contribution to the $++---$ amplitude is
absent, as there is no $++-$ vertex.} {\epsfbox{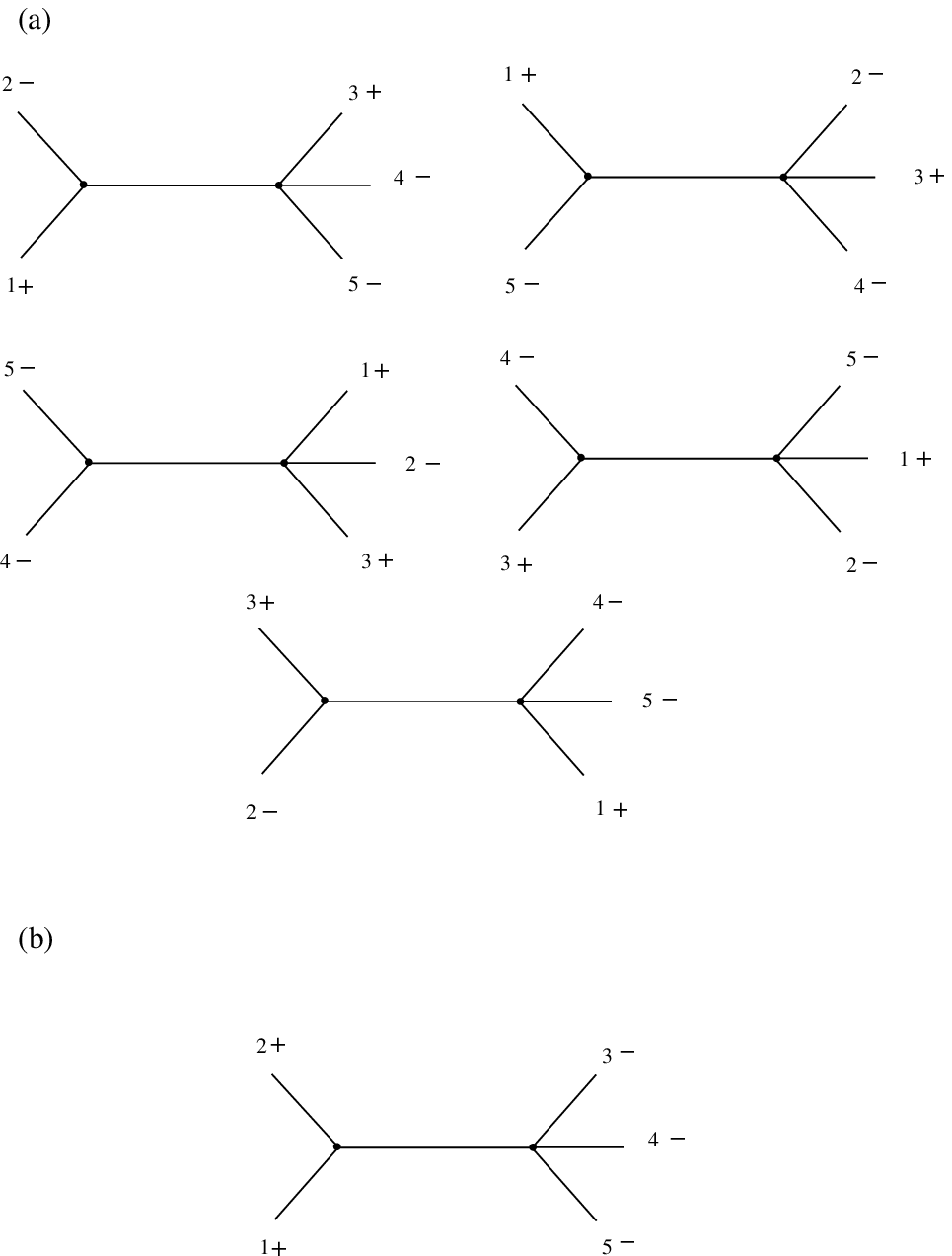}}

The next case is the five gluon amplitude with two gluons of
positive helicity and three of negative helicity.  In general,
five MHV tree diagrams contribute, as sketched in figure 3a for
the case of helicities $+-+--$.  Each diagram contains two
vertices, one of them trivalent and one four-valent.  The vertices
are all defined off-shell by the same procedure as above.  The sum
of the five MHV tree diagrams can be shown, with the aid of
symbolic manipulation, to coincide with the standard tree level
amplitude for this process, which is \eqn\sonsonno{{[1~3]^4\over
[1~2][2~3][3~4][4~5][5~1]}.} For the helicity configuration
$++---$, there are only four MHV tree diagrams; there is no
contribution of the form sketched in figure 3b, since by
definition each vertex in an MHV tree diagram absorbs precisely
two gluon lines of negative helicity.  We have  verified with the
help of symbolic manipulation that the sum of the four remaining
diagrams reproduces the standard result (which is obtained from
\sonsonno\ by simply replacing $[1~3]^4$ in the numerator by
$[1~2]^4$).

We have made similar verifications in a number of additional
cases, including five gluon amplitudes with helicity $+----$, all
the six gluon amplitudes with three or four gluons of negative
helicity, all the seven gluon amplitudes with three gluons of
negative helicity, and finally the $++-----$ amplitude.

\ifig\teramp{MHV tree diagrams contributing to the $---++\dots +$
amplitude.} {\epsfbox{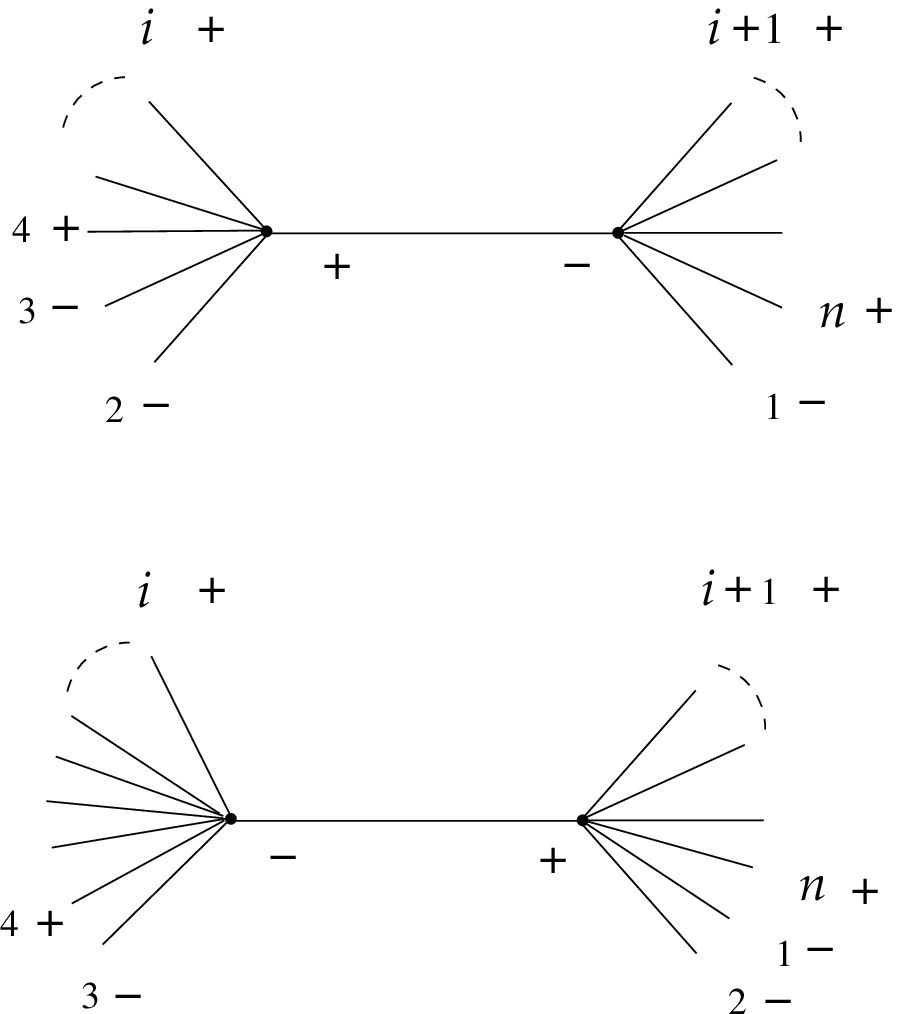}}

Our claim that conventional Yang-Mills tree amplitudes coincide
with the amplitudes computed from MHV tree diagrams implies
surprisingly simple formulas for some amplitudes.  For example,
consider $n$ gluon amplitudes with precisely three gluons of
negative helicity -- the next case after the simple MHV
amplitudes.  They come from MHV tree diagrams with precisely two
vertices and one propagator.  If the helicities are $---+++\dots
++$, i.e., the three gluons of negative helicity are consecutive
(see \kosower\ for a previous evaluation of these amplitudes),
there are precisely $2(n-3)$ possible diagrams, sketched in figure
4. They can be evaluated to give
\eqn\mmmM{\eqalign{A&=\sum_{i=3}^{n-1} {\vev{1~ (2,i)}^3 \over
\vev{(2,i)~i+1}\vev{i+1~i+2} \dots \vev{n~1}} {1\over
q_{2i}^2}{\vev{2~3}^3 \over \vev{(2,i)~2} \vev{3~4}\dots \vev{i
~(2,i)}}\cr&+\sum_{i=4}^n {\vev{1~2}^3\over \vev{2~(3,i)}
\vev{(3,i)~i+1} \dots \vev{n~1}} {1\over q_{3i}^2}
{\vev{(3,i)~3}^3 \over  \vev{3~4} \dots \langle
i-1~i\rangle\vev{i~(3,i)}}.}} Here we define
$q_{ij}=p_i+p_{i+1}+\dots+p_j$; the corresponding spinor
$\lambda_{ij\,a}$ is defined in the usual way as
$\lambda_{ij\,a}=q_{ij\,a\dot a}\eta^{\dot a}$, and
$\vev{i~(j,k)}$ is an abbreviation for
$\vev{\lambda_i,\lambda_{jk}}$.

For other orderings of the external helicities, the number of
diagrams is greater, but grows for large $n$ at most as $n^2$. If
$S,T,$ and $U$ are the number of positive helicity gluons between
successive gluons of negative helicity (so $S+T+U=n-3$), then the
number of diagrams is $2 ( n - 3 )  + S T + T U + U S $.

The formula \mmmM\ is not manifestly covariant in general, but it
becomes so if we pick $\eta^{\dot a}$ to equal one of the
$\tilde\lambda_i^{\dot a}$. (We show in section 5 that the
amplitude is independent of the choice of $\eta^{\dot a}$.) If
$\eta^{\dot a}=\tilde\lambda_2^{\dot a}$,  the amplitude becomes
\eqn\simm{ \eqalign{ A =& { 1 \over \prod_{k=3}^n
\vev{k~k+1}}\left[ \sum_{i=4}^{n-1} {\vev{i~i+1}\over
\vev{i^-|{q\!\!\!/}_{2,i}|2^-}
\vev{(i+1)^-|{q\!\!\!/}_{i+1,2}|2^-}
\vev{2^-|{q\!\!\!/}_{2,i}|2^-}} \right. \cr & \left. \left(
{\vev{3~2}^3 \vev{1^-|{q\!\!\!/}_{2,i}|2^-}^3\over q_{2,i}^2} +
{\vev{1~2}^3 \vev{3^-|{q\!\!\!/}_{i+1,2}|2^-}^3\over q_{i+1,2}^2}
\right) + A_{3,n}
 \right]  }}
where we have introduced the manifestly Lorentz covariant notation
$\vev{ m^-|{p\!\!\!/}\;\;|r^-} =m^ap_{a\dot a}r^{\dot a}$ and used
the fact that $q_{3,i}=-q_{i+1,2}$. $A_{3,n}$ is the contribution
from the $i=3$ and $i=n$ terms of the first and second sums in
(3.7) respectively. We have to treat these terms separately,
because they have a factor of $[2~\eta]$ in the denominator, which
vanishes for $\eta^{\dot a} = \tilde\lambda^{\dot a}_2$. However,
combining them and using Schouten's identity\foot{This identity
asserts that for any four spinors $\alpha,\beta,\gamma,\delta$, we
have
$\vev{\alpha,\beta}\vev{\gamma,\delta}+\vev{\alpha,\gamma}\vev{\delta,\beta}
+\vev{\alpha,\delta}\vev{\beta,\gamma}=0$.} one finds a factor of
$[2~\eta]$ in the numerator as well. Thus, the substitution
$\eta^{\dot a} = \tilde\lambda^{\dot a}_2$ can be made to get
\eqn\twott{ A_{3,n} = -\vev{1~3}^2 \left( {s_{13}+ 2 (s_{12} +
s_{23})\over [3~2][1~2]} + {\vev{1~2}\vev{n~3}\over
[1~2]\vev{n~1}} + {\vev{3~2}\vev{1~4}\over [3~2]\vev{3~4}}
\right)}
where $s_{km}=(p_k+p_m)^2 = \vev{k~m}[k~m]$.

The amplitude \simm\ is manifestly Lorentz-covariant and
bose-symmetric. Bose symmetry merely says that the amplitude
should be invariant under rotations and reflections of the chain
that preserve the helicities.  For this particular amplitude, the
only such symmetry is the reflection that maps particle $k$ to
particle $4-k$; we have chosen $\eta$ in a way that preserves this
symmetry.  One could also obtain different but manifestly
Lorentz-covariant and bose-symmetric expressions for the same
amplitude by averaging over the choices $\eta^{\dot
a}=\tilde\lambda_k^{\dot a}$ and $\eta^{\dot
a}=\tilde\lambda_{4-k}^{\dot a}$, for some fixed $k$. (Because of
Schouten's identity, the various spinor products are not
independent, and quite different-looking formulas can be written
for the same amplitudes.)  We have verified that \simm\ agrees
with the standard result for the $---+++$ amplitude.

\newsec{Collinear And Multiparticle Singularities}

Here we will show that MHV tree graphs generate amplitudes with
the correct collinear and multiparticle singularities.

Collinear singularities arise when (for example) the momenta of
two incoming particles in a scattering amplitude are proportional,
so that their sum is also lightlike.

We describe a collinear singularity as a process with two
particles going to one, so we consider, for example, the collinear
singularity $++\to +$ with two initial gluons of positive helicity
combining to one of positive helicity.  Since crossing symmetry
reverses the helicity of a gluon, and our convention for vertices
in a Feynman diagram is to consider all gluons incoming, the
$++\to +$ collinear singularity receives a contribution from a
$++-$ interaction vertex.

In MHV tree diagrams, a singularity when two gluons $i$ and $i+1$
develop collinear momenta will only arise if these two gluons are
attached to the same vertex in the graph. In Yang-Mills theory,
the collinear singularities are $++\to +$, $+-\to -$,   $+-\to +$,
and $--\to -$.  (There are no $++\to -$ or $--\to +$ collinear
singularities, since there are no $+++$ or $---$ interaction
vertices.)  For our purposes, there are really two kinds of
collinear singularity: (a) for $++\to +$ and $+-\to -$, the number
of negative helicity gluons is conserved; (b) for $--\to -$ and
$+-\to +$, the number of negative helicity gluons is reduced by
one.

In case (a), the limit as gluons $i$ and $i+1$ become collinear is
extracted by merely taking the collinear limit of the MHV vertex
to which they are attached.  This MHV vertex is a standard
Yang-Mills scattering amplitude and has the standard collinear
singularities.  (Our off-shell continuation of an MHV  vertex is
easily seen not to modify the collinear singularities for the
on-shell particles in that vertex.) Thus, it is manifest that MHV
tree diagrams have the correct collinear singularities of type
(a).

For singularities of type (b), we need only to be a little more
careful.  Consider an  MHV tree diagram in which gluons $i$ and
$i+1$, of helicities $--$ or $+-$, are attached to a vertex with
$k$ gluons (some of which may be off-shell),
for some $k\geq 3$.  If $k\geq 4$, this diagram will
not contribute any collinear singularity of type (b), since the
``mostly plus'' MHV tree amplitudes do not have any $--\to -$ or
$+-\to +$ collinear singularities.

\ifig\triamp{Diagrams contributing to collinear singularities of
type (b).  The shaded ``blob'' represents the complete $n-1$ gluon tree
amplitude.} {\epsfbox{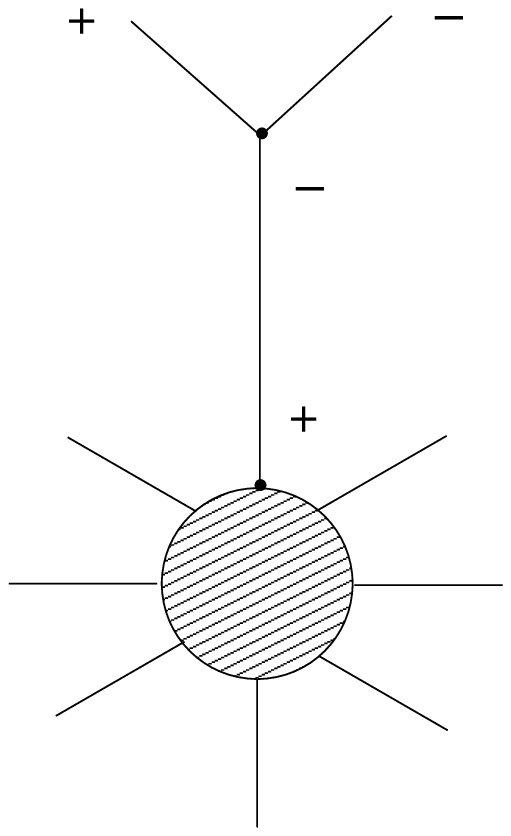}} The collinear singularities of
type (b) will therefore come entirely from diagrams with $k=3$, as
in figure 5.  In the collinear limit, $P$ becomes on-shell, and
the spinor $\lambda_P$ as we have defined it (namely
$\lambda_{P\,a}=P_{a\dot a}\eta^{\dot a}$) becomes a multiple of
the spinor arising in the factorization $P_{a\dot
a}=\lambda_a\tilde\lambda_{\dot a}$.  A rescaling of $\lambda_P$
does not matter.  So assuming inductively that the MHV tree
diagrams give correctly the $n-1$ gluon tree amplitude (represented by the
``blob'' in figure 5), the configuration of figure 5
manifestly gives the correct type (b) collinear singularity.

One can see in a similar fashion that our recipe reproduces the
correct multiparticle poles in tree amplitudes.  Consider an
$n$-particle amplitude and pick some $i$ and $j$ such that the set
of particles $i,i+1,\dots, j$ and the set $j+1,j+2,\dots,i-1 $
each have at least three elements.
  Let $P=p_i+p_{i+1}+\dots +p_j$.
The multiparticle singularity in this channel has an amplitude
that is simply $1/P^2$ times the product of the tree amplitudes in
the subchannels. It arises in our formalism from MHV tree diagrams
(figure 6) with a single offshell gluon of momentum $P$ connecting
the two clusters.  The propagator of the off-shell gluon is
$1/P^2$, and as $P^2\to 0$, the spinor $\lambda_P$ of this gluon,
as we have defined it, becomes the standard spinor of an on-shell
gluon of momentum $P$. So assuming inductively that MHV tree
graphs describe the lower order amplitudes correctly, they
describe the correct multiparticle poles in the $n$ gluon
amplitude.

\ifig\triramp{MHV tree diagrams with multiparticle singularities
in a particular channel that carries momentum $P$.  The shaded
``blobs'' represent subamplitudes computed with MHV tree
diagrams.} {\epsfbox{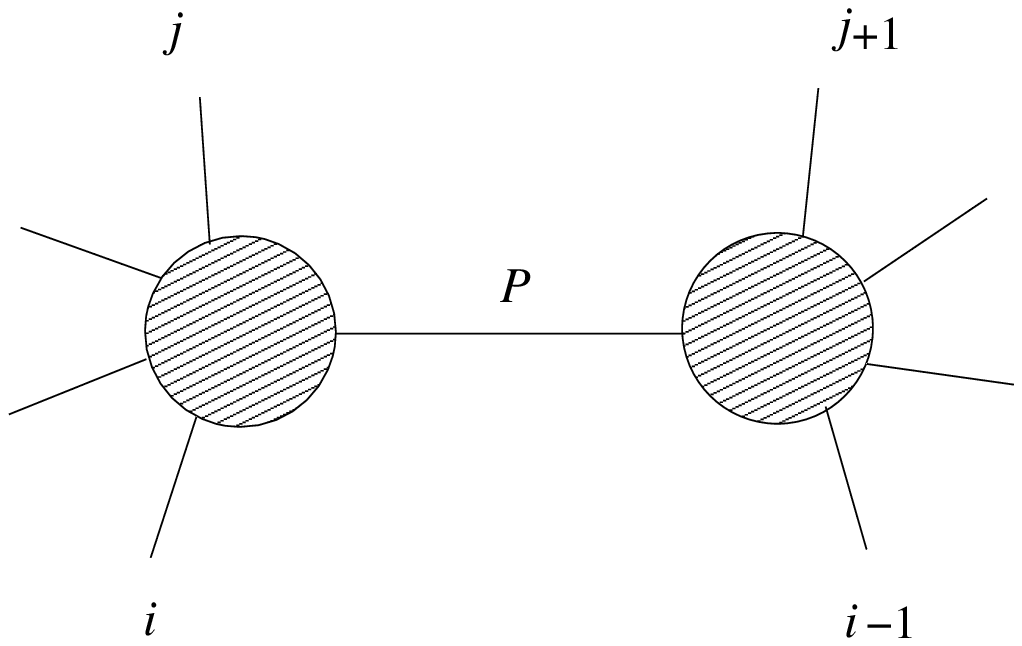}}

\newsec{Covariance Of The Amplitudes}

Here we will demonstrate that the sum of MHV tree amplitudes is
Lorentz covariant.  For simplicity, we consider the case of
diagrams with only one propagator (and therefore precisely three
external gluons of negative helicity), but we do not believe that
this restriction is essential. We present the argument here
without relation to twistor theory, because the covariance of the
sum of MHV trees is of interest irrespective of any connection to
twistors. However, the argument was suggested by a nonrigorous
twistor analysis that we present in the next section.

Consider as in figure 7 an $n$-gluon tree diagram with one
propagator.  The external gluons are divided into two sets $L$ and
$R$ of gluons attached to the left or right in the diagram; the
internal line carries a momentum $P=\sum_{i\in L}p_i$. We have no
natural way to assign spinors $\lambda$, $\tilde\lambda$ to the
internal line (since in general $P^2\not= 0$), so instead we
introduce an arbitrary $\lambda$ and $\tilde\lambda$ associated
with this line; we will integrate over $\lambda$ and
$\tilde\lambda$ in a manner that will be described.

 \ifig\trampeto{\vskip .5 cm \noindent MHV diagrams
with two vertices, labeled $L$ and $R$, connected by a propagator
that carries momentum $P$.} {\epsfbox{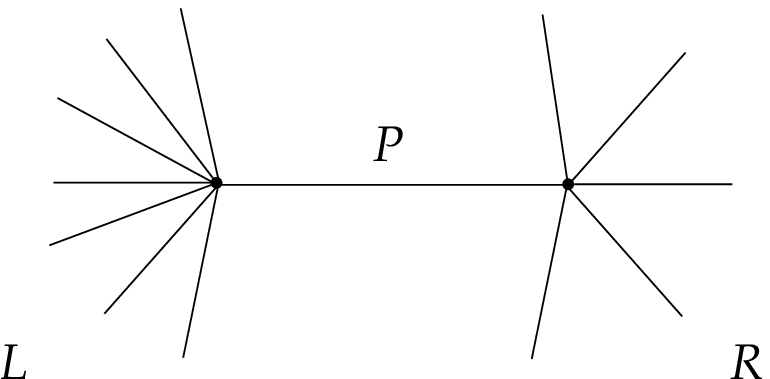}} The gluons
attached on the left vertex of figure 7 make up a set $L'$
consisting of $L$ plus the internal gluon, and similarly the
gluons on the right make up a set $R'$ consisting of $R$ plus the
internal gluon. $L'$ and $R'$ each comes with a natural cyclic
order. In an MHV tree diagram, the amplitudes at the left and
right vertices are \eqn\uvur{\eqalign{g_L(\lambda_i\vert_{i\in
L'})={\langle\lambda_{x_L},\lambda_{y_L}\rangle^4\over
 \prod_{i\in L'}\langle\lambda_i,\lambda_{i+1}\rangle}\cr
g_R(\lambda_i\vert_{i\in
R'})={\langle\lambda_{x_R},\lambda_{y_R}\rangle^4\over
 \prod_{i\in R'}\langle\lambda_i,\lambda_{i+1}\rangle}.\cr}} $x_L$ and
 $y_L$ are the labels of the negative helicity gluons on the left,
 and $x_R, $ $y_R$ are the analogous labels on the right.
 The dependence of $g=g_Lg_R$ on $\lambda$ is
extremely simple:
\eqn\orfon{g={\langle\lambda_\sigma,\lambda\rangle^4\over
\prod_{\alpha=1}^4\langle\lambda_\alpha,\lambda\rangle}\tilde g,}
where $\tilde g$ is independent of $\lambda$.   Here two poles in
the denominator come from $g_L$ and two from $g_R$; $\alpha$ runs
over the four gluons that in the cyclic order are adjacent to the
internal line  on either the left or the right. $\sigma$ is the
negative chirality gluon on the same side ($L$ or $R$) on which
the internal line carries negative helicity.  In particular, $g$
is invariant under scalings of $\lambda$.

Now we write down the integral that we will consider: \eqn\torog{
I_\Gamma={i\over 2\pi}\int \langle\lambda,d\lambda\rangle
[\tilde\lambda,d\tilde\lambda]{1\over (P_{a\dot
a}\lambda^a\tilde\lambda^{\dot a})^2} g(\lambda;\lambda_i).} The
integration ``contour'' is described momentarily. We call this
integral $I_\Gamma$ to emphasize the fact that it depends on the
choice of a particular MHV tree graph $\Gamma$. Since $g$ is
invariant under scalings of $\lambda$ or $\tilde\lambda$ (and in
fact is independent of $\tilde\lambda$), the integrand in \torog\ is also
invariant under this scaling and
 makes sense as a meromorphic two-form on $\Bbb{CP}^1\times
\Bbb{CP}^1$.  Here the $\lambda^a$ are homogeneous coordinates on
one $\Bbb{CP}^1$, and $\tilde\lambda^{\dot a}$ on the second
$\Bbb{CP}^1$.

When we actually evaluate the integral, we will take the
integration ``contour'' to be a two-sphere $S$ defined by saying
that $\tilde\lambda$ is the complex conjugate of $\lambda$. This
ensures that the vector $w_{a\dot a}=\lambda_a\tilde\lambda_{\dot
a}$ is real, nonzero, and lightlike.  It follows that if $P$ is
real and  timelike, the denominator $(P_{a\dot a}w^{a\dot a})^2$
in the definition of $I_\Gamma$ is everywhere nonzero.  The only
singularities of the integrand are the simple poles of $g$, which
do not affect the convergence of the integral.  The integral over
$S$ is hence convergent for timelike $P$. We use the integral to
define $I_\Gamma$ as an analytic function of $P$ (and the other
variables) which can then be continued beyond the real, timelike
region. In fact, our evaluation of the integral will give such a
continuation.

 For the moment,
however, we continue algebraically without interpreting
$\tilde\lambda$ as the complex conjugate of $\lambda$.  As in the
definition of MHV tree diagrams, we introduce an arbitrary spinor
$\eta^{\dot a}$ of negative chirality, and we find the identity
\eqn\ronron{ {[\tilde\lambda,d\tilde\lambda]\over (P_{a\dot
a}\lambda^a\tilde\lambda^{\dot a})^2}=-d\tilde\lambda^{\dot c}
{\partial\over\partial\tilde\lambda^{\dot
c}}\left([\tilde\lambda,\eta]\over (P_{a\dot
a}\lambda^a\tilde\lambda^{\dot a})(P_{b\dot b}\lambda^b\eta^{\dot
b})\right).} Since $g$ is independent of $\tilde\lambda$, it
trivially follows that likewise \eqn\konron{
{[\tilde\lambda,d\tilde\lambda]\,g(\lambda;\lambda_i)\over
(P_{a\dot a}\lambda^a\tilde\lambda^{\dot
a})^2}=-d\tilde\lambda^{\dot c}
{\partial\over\partial\tilde\lambda^{\dot
c}}\left([\tilde\lambda,\eta]\,g(\lambda;\lambda_i)\over (P_{a\dot
a}\lambda^a\tilde\lambda^{\dot a})(P_{b\dot b}\lambda^b\eta^{\dot
b})\right).}

At this point, we interpret $\tilde\lambda$ as the complex
conjugate of $\lambda$.  If $\lambda^a=(1,z)$, then
$\tilde\lambda^{\dot a}=(1,\bar z)$; the integration region $S$ is
the complex $z$ plane including a point at infinity.   The
operator $d\tilde\lambda^{\dot
a}\partial/\partial\tilde\lambda^{\dot a}$ is $d\bar
z(\partial/\partial\bar z)$, and if \konron\ were precisely true,
it would follow upon integration by parts that $I_\Gamma$ is
identically zero. Actually, once we interpret $\tilde\lambda$ as
the complex conjugate of $\lambda$, the formula acquires delta
function contributions, since \eqn\yuro{{\partial\over
\partial\bar z}{1\over z-b} =2\pi\delta(z-b).}
The delta function is normalized so that $\int |dz\,d\bar
z|\,\delta(z-b)=1$. This also means that in terms of differential
forms, $\int dz\wedge d\bar z\, \delta(z-b)=-i=-\int d\bar z\wedge
dz\,\delta(z-b)$, since if $z=x+iy$ with $x,y$ real, then
$dz\wedge d\bar z=-2i dx\wedge dy=-i|dz\,d\bar z|$.  It is also
convenient to write $\bar\delta(z-b)=\delta(z-b)d\bar z$, and more
generally, for any holomorphic function $f$,
\eqn\jonon{\bar\delta(f)=\delta(f)d\bar f.} (Thus $\bar\delta(f)$
is a $\bar\partial$-closed $(0,1)$-form, a property that we will
use in section 6.) So \eqn\bonon{\int dz\,\bar\delta(z-b)=-i.} We
can write \yuro\ in a more covariant form:
\eqn\onso{d\tilde\lambda^{\dot
c}{\partial\over\partial\tilde\lambda^{\dot c}} {1\over
\langle\zeta,\lambda\rangle}=2\pi\bar\delta(\langle\zeta,\lambda\rangle),}
again assuming $\tilde\lambda=\bar\lambda$. The idea here is that
in coordinates with $\lambda^a=(1,z)$, $\tilde\lambda^{\dot a}
=(1,\bar z)$, $\zeta^a=(1,b)$, \onso\ reduces to \yuro. If
$\lambda^a=(1,z)$, then $\vev{\lambda,d\lambda}=dz$, so a more
covariant version of \bonon\ is the statement that if $B(\lambda)$
is any function that is homogeneous of degree $-1$, then
\eqn\sxo{\int \langle\lambda,d\lambda\rangle
\bar\delta(\vev{\zeta,\lambda}) B(\lambda) =-iB(\zeta).}

In evaluating
\konron\ more precisely to include such delta functions, we need
not be concerned about singularities from zeroes of $P_{a\dot
a}\lambda^a\tilde\lambda^{\dot a}$, since as we have discussed,
this function has no zeroes in the integration region.  However,
we get a contribution that we will call $I_{\Gamma,\eta}$ from the
pole at \eqn\rorno{\lambda_a=P_{a\dot a}\eta^{\dot a}}
 that comes from the vanishing of the factor $P_{a\dot a}\lambda^a\eta^{\dot
a}$ in the denominator. And we get four contributions that we will
call $I_{\Gamma,\alpha}$ from the poles at
$\lambda=\lambda_\alpha$ which are visible in the formula \orfon\
for $g$.  The condition \rorno\ should be familiar; it was used in
section 2 to make an off-shell continuation of the MHV amplitudes.

\ifig\trampop{The graphs contributing a pole at
$\lambda=\lambda_\alpha$.  Each vertex has a natural cyclic order,
which we take to be counterclockwise, as indicated by the arrows.
 In one graph, $\alpha$ is on the left, just ahead of the internal line,
 and in the other graph, it is on the right, just after
it. The reversed order reverses
 the sign of the residue of the pole.} {\epsfbox{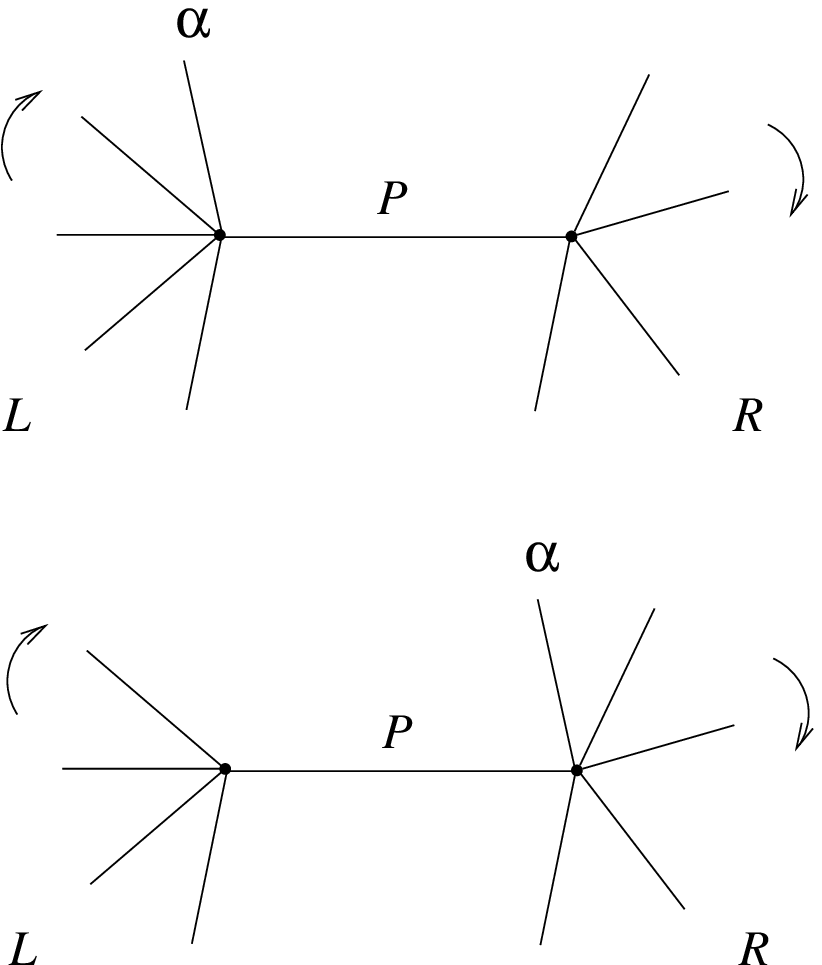}}

So we can schematically write
\eqn\yorfon{I_\Gamma=I_{\Gamma,\eta}+\sum_\alpha
I_{\Gamma,\alpha}.} To evaluate $I_{\Gamma,\eta}$, and
$I_{\Gamma,\alpha}$, we evaluate \konron\ more precisely,
including the delta functions that should be included when
$\tilde\lambda$ is understood as the complex conjugate of
$\lambda$. We have \eqn\kononron{\eqalign{
{[\tilde\lambda,d\tilde\lambda]\,g(\lambda;\lambda_i)\over
(P_{a\dot a}\lambda^a\tilde\lambda^{\dot
a})^2}=&-d\tilde\lambda^{\dot c}
{\partial\over\partial\tilde\lambda^{\dot
c}}\left([\tilde\lambda,\eta]\,g(\lambda;\lambda_i)\over (P_{a\dot
a}\lambda^a\tilde\lambda^{\dot a})(P_{b\dot b}\lambda^b\eta^{\dot
b})\right)\cr &+{2\pi [\tilde\lambda,\eta]\over P_{a\dot
a}\lambda^a \tilde\lambda^{\dot a}} \left(-\bar\delta(P_{b\dot
b}\lambda^b\eta^{\dot b})g+{1\over P_{b\dot b}\lambda^b\eta^{\dot
b}}\sum_{\alpha=1}^4 \bar\delta(\langle\lambda_\alpha, \lambda\rangle)
{\langle\lambda_\sigma, \lambda_\alpha\rangle^4\over
\prod_{\beta\not=\alpha}\langle\lambda_\beta,\lambda
_\alpha\rangle}\tilde g\right).\cr}}

We can now evaluate $I_{\Gamma,\eta}$, which is the contribution
of the delta function that is supported at $\lambda_a=P_{a\dot
a}\eta^{\dot a}$. At $\lambda_a=P_{a\dot a}\eta^{\dot a}$, we have
$ [\tilde\lambda,\eta]/ P_{a\dot a}\lambda^a \tilde\lambda^{\dot
a}=-1/(\half P_{a\dot a}P^{a\dot a})=-1/P^2$.
So \eqn\ormigo{I_{\Gamma,\eta}= {1\over
P^2}g(\lambda_P;\lambda_i),} where as in section 2,
$\lambda_{P\,a}=P_{a\dot a}\eta^{\dot a}$. In other words,
$I_{\Gamma,\eta}$ is simply the amplitude, as defined in section
2, for the MHV tree graph $\Gamma$. Similarly,
\eqn\forno{I_{\Gamma,\alpha}= {2\pi [\tilde\lambda_\alpha,\eta]\over
(P_{a\dot a} \lambda_\alpha^a \tilde\lambda_\alpha^{\dot
a})(P_{b\dot b}\lambda_\alpha^b\eta^{\dot b})}
 {\langle\lambda_\sigma,
\lambda_\alpha\rangle^4\over
\prod_{\beta\not=\alpha}\langle\lambda_\beta,\lambda
_\alpha\rangle}\tilde g= {2\pi [\tilde\lambda,\eta]\over (P_{a\dot
a} \lambda_\alpha^a \tilde\lambda_\alpha^{\dot a})(P_{b\dot
b}\lambda_\alpha^b\eta^{\dot b})}{\rm
Res}_{\lambda=\lambda_\alpha}\,g(\lambda;\lambda_i).}

Upon summing over all tree graphs with the given set of external
gluons, we have \eqn\porfon{ \sum_\Gamma I_{\Gamma}=\sum_\Gamma
I_{\Gamma,\eta}+\sum_\alpha I_{\Gamma,\alpha}.} We will see
shortly that \eqn\burgog{\sum_\Gamma I_{\Gamma,\alpha}=0}  for all
$\alpha$. Given this, we have \eqn\irfon{\sum_\Gamma I_\Gamma=
\sum_\Gamma I_{\Gamma,\eta}.} Since the left hand side is Lorentz
covariant (a statement that we explain more fully below), it
follows as we have promised that the sum of MHV tree amplitudes is
covariant.

 Now we will verify \burgog. We consider two graphs $\Gamma_1$ and
 $\Gamma_2$ -- selected as in figure 8 -- for
 which the function $g$ has a pole at $\lambda=\lambda_\alpha$.
They differ by whether the gluon $\alpha$ is in $L$, just before
the internal gluon (in the cyclic order), or in $R$, just after
it.  Because of this difference in ordering, when we evaluate
$g=g_Lg_R$ using \uvur, one $g$ function contains
 a factor $1/\langle\lambda,\lambda_\alpha\rangle$ while the other contains
a factor $1/\langle\lambda_\alpha,\lambda\rangle$.  The other
factors in the two $g$ functions, which we will call $g_1$ and
$g_2$, become equal when we set $\lambda=\lambda_\alpha$. So ${\rm
Res}_{\lambda=\lambda_\alpha} \,g_{1}=- {\rm
Res}_{\lambda=\lambda_\alpha} \,g_{2}$. The other factor in
\forno\ that we must consider in comparing $I_{\Gamma_1,\alpha}$
and $I_{\Gamma_2,\alpha}$ is $X=1/(P_{a\dot a} \lambda_\alpha^a
\tilde\lambda_\alpha^{\dot a})(P_{b\dot
b}\lambda_\alpha^b\eta^{\dot b})$. The two graphs have different
$P$'s, but as the $P$'s differ by $P_{a\dot a}\to P_{a\dot
a}+\lambda_{\alpha\,a}\tilde\lambda_{\alpha\,\dot a}$, they
 have the same value of $X$.  So finally, the two graphs give equal and opposite
 poles at $\lambda=\lambda_\alpha$.  All poles at
 $\lambda=\lambda_\alpha$ are canceled in this way among pairs of
 graphs.

\bigskip\noindent{\it A Subtle Detail}

There is actually one further subtlety in this argument (which
some readers may wish to omit). Suppose that on the left of the
first diagram in figure 8 there are only two external gluons --
one labeled $\alpha$ and one labeled, say, $\beta$. The evaluation
of the diagram as above yields a pole at $\lambda=\lambda_\alpha$
that must be canceled by a similar pole when $\alpha$ has moved to
the right. In that contribution,  only one gluon, namely $\beta$,
remains on the left (figure 9). We therefore have to allow
contributions in the present analysis in which only two gluons
(one of them off-shell) are attached to the vertex on the left.
This presents a riddle, since the MHV tree diagrams have no such
divalent vertices.

Let us see examine this more closely.   In figure 9, both $\beta$
and the internal gluon joining to $L$ have negative helicity
(since they are the only candidates for the two negative helicity
gluons on $L$).  Hence in \orfon, $\sigma$ and two of the
$\alpha$'s are both equal to $\beta$, so $g$ becomes
\eqn\ucu{g=\vev{\lambda_\beta,\lambda}^2\prod_{\nu}{1\over
{\vev{\lambda_\nu,\lambda}}}\tilde g,} where $\nu$ runs over the
two neighbors of the internal gluon in $R$. In particular, there
is no pole at $\lambda=\lambda_\beta$ (and so no need to cancel
its residue by introducing a contribution with only one gluon
attached to $L$).

\ifig\tramppok{A diagram with a divalent vertex on the left.  The
two gluons entering the vertex both have negative helicity.  The
external gluon is labeled $\beta$, and its momentum $p_\beta$ also
equals the momentum $P$ of the internal gluon.}
{\epsfbox{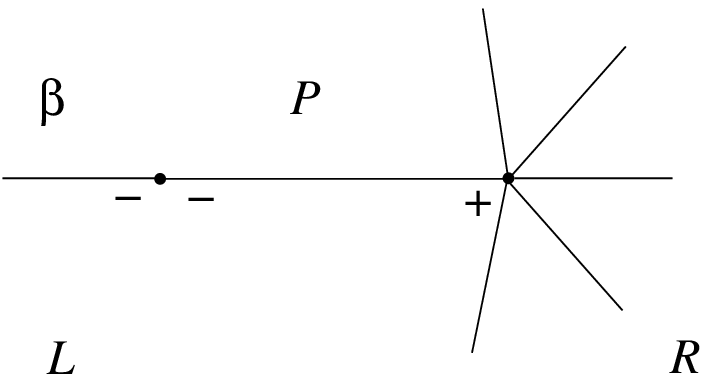}}

 Since $P=p_\beta$, we
have $P_{a\dot a}=\lambda_{\beta\,a}\tilde\lambda_{\beta\,\dot
a}$.  The integral representation of $I_\Gamma$ becomes
\eqn\nnos{I_\Gamma={i\over
2\pi}\int\vev{\lambda,d\lambda}[\tilde\lambda,d\tilde\lambda]
{1\over [\tilde\lambda,\tilde\lambda_\beta]^2} \prod_{\nu}{1\over
\vev{\lambda_\nu,\lambda}} \tilde g,} where a factor of
$\vev{\lambda,\lambda_\beta}^2$ in the denominator has canceled
such a factor in the numerator of \ucu.  This cancellation ensures
that the integral for $I_\Gamma$ is convergent (if we integrate
symmetrically near  $\tilde\lambda=\tilde\lambda_\beta$ where the
denominator has its strongest singularity) even though $P$ is
lightlike.

We can evaluate the integral by a sum of residues.
 First consider the contribution
$I_{\Gamma,\nu}$ from the pole at $\lambda=\lambda_\nu$ (for
either of the two possible values of $\nu$). By \forno, it is
\eqn\honno{I_{\Gamma,\nu}={2\pi [\tilde\lambda_\nu,\eta]\over
(P_{a\dot a}\lambda_{\nu}^a\tilde\lambda_{\nu}^{\dot a})(P_{b\dot
b}\lambda_\nu^b\eta^{\dot b})}{\rm
Res}_{\lambda=\lambda_\nu}g(\lambda;\lambda_i).} With only
one gluon on $L$, we have $P=p_\beta$.  So $P_{b\dot
b}=\lambda_{\beta\,b}\tilde\lambda_{\beta\,\dot b}$, whence
\eqn\fonno{I_{\Gamma,\nu}={2\pi[\tilde\lambda_\nu,\eta]\over
\vev{\lambda_\beta,\lambda_\nu}^2[\tilde\lambda_\beta,\tilde\lambda_\nu]
[\tilde\lambda_\beta,\eta]} {\rm
Res}_{\lambda=\lambda_\nu}g(\lambda;\lambda_i).}

From \ucu, if we write $\nu_i$, $i=1,2$ for the two possible
values of $\nu$, this gives
\eqn\jonno{I_{\Gamma,\nu_i}={2\pi[\tilde\lambda_{\nu_i},\eta]\over
[\tilde\lambda_\beta,\tilde\lambda_{\nu_i}][\tilde\lambda_\beta,\eta]}
{\tilde g\over \vev{\lambda_{\nu_{i'}},\lambda_{\nu_i}}},} where
$\nu_{i'}\not= \nu_i$.  From this it follows (using the Schouten
identity  of footnote 2 to combine the terms) that
$\sum_{i=1,2}I_{\Gamma,\nu_i}$ is independent of $\eta$. Hence,
unlike the cases with more than two gluons attached to $L$, we do
not have to add an additional contribution from a pole at
$\lambda_a=P_{a\dot a}\eta^{\dot a}$ to cancel the
$\eta$-dependence.

We do not want such a contribution, since, with the vertex on the
left of figure 9 being divalent, it does not correspond to
anything in the MHV tree diagrams of sections 2 and 3. In more
general cases with a $k$-valent vertex of $k\geq 3$, the
contribution that we called $I_{\Gamma,\eta}$ arises from the
singularity of \eqn\yoyo{{2\pi
[\tilde\lambda,\eta]g(\lambda;\lambda_i)\over(P_{b\dot
b}\lambda^b\tilde\lambda^{\dot b})(P_{a\dot a}\lambda^a\eta^{\dot
a})}} at $P_{a\dot a}\lambda^a\eta^{\dot a}=0$. With $P_{a\dot
a}=\lambda_{\beta\,a}\tilde\lambda_{\beta\,\dot a}$, this
singularity would be at $\vev{\lambda_\beta,\lambda}=0$, but in
\yoyo, there is no singularity there, because $g$ is divisible by
$\vev{\lambda_\beta,\lambda}^2$. Thus, configurations with a
divalent vertex
 have a nonvanishing  $I_{\Gamma,\alpha}$ and participate in the associated
cancellation, but have vanishing $I_{\Gamma,\eta}$ and do
not contribute to the MHV tree diagrams.

\bigskip\noindent{\it Covariance Of The Amplitude}

Finally, the assertion that $I=\sum_\Gamma I_\Gamma$ is Lorentz
covariant needs some elaboration:

(1) The integral representation \torog\ appears to show that
$I_\Gamma $ is holomorphic in the $\lambda_i$  and in the
$\tilde\lambda_i$ (the latter enter only via $P$). Though the
holomorphy in $\tilde\lambda_i$ is valid, the holomorphy in
$\lambda_i$ fails because of the poles: the $\bar\partial$
operator of $\lambda_\alpha$, namely $d\bar\lambda_\alpha^a
\,\partial/\partial\bar\lambda_\alpha^a$, in acting on the
integrand of $I_\Gamma$, produces a delta function at
$\lambda=\lambda_\alpha$. When we write
$I_\Gamma=I_{\Gamma,\eta}+\sum_\alpha I_{\Gamma,\alpha}$, the
first term $I_{\Gamma,\eta}$ is holomorphic in the
$\lambda_\alpha$, but the $I_{\Gamma,\alpha}$ are not.

(2) The integral  \torog\ defining $I_\Gamma$ formally has
$SL(2,\Bbb{C})\times SL(2,\Bbb{C})$ symmetry, where one $SL(2)$
acts on spinor indices $a,b$ and the other on spinor indices $\dot
a,\dot b$.  Thus, one $SL(2)$ acts on $\lambda,\lambda_i,$ and the
other on $\tilde\lambda,\tilde\lambda_i$. $SL(2)\times SL(2)$ is a
double cover of the complexified Lorentz group.

(3) The choice of integration contour $S$ with
$\tilde\lambda=\bar\lambda$ breaks $SL(2)\times SL(2)$ down to the
diagonal $SL(2)$, which is a double cover of the real Lorentz
group $SO(3,1)$. Were there no poles, a contour deformation
argument would show that the integral possesses the full
$SL(2)\times SL(2)$ symmetry, even though the contour does not.
Because of the poles, the full $SL(2)\times SL(2)$ invariance is
not restored upon doing the integral and $I_\Gamma$ is only
invariant under the diagonal $SL(2)$.

(4) After summing over $\Gamma$, the $I_{\Gamma,\alpha}$ cancel,
as we argued above, and hence holomorphy in the $\lambda_i$ is
restored.

(5) The sum $I=\sum_\Gamma I_\Gamma=\sum_\Gamma I_{\Gamma,\eta}$
is accordingly holomorphic in the $\lambda_i$ and
$\tilde\lambda_i$.  The real Lorentz group, or rather its double
cover $SL(2)$, acts holomorphically on these variables leaving $I$
invariant, and hence $I$ is automatically invariant under the
complexification of this group, which is the full $SL(2)\times
SL(2)$.

\newsec{Heuristic Analysis Of Disconnected Twistor Diagrams}

Here we will make a nonrigorous analysis of the disconnected
twistor diagrams that contribute to the amplitudes studied in the
last section.  Interpreting the interaction vertices in the
Feynman diagram of figure 7 as  degree one instantons in twistor
space, and the line connecting the vertices as a twistor
propagator, we will explain what manipulations applied to this
twistor configuration give the integral studied in the last
section.

We are going to use somewhat different twistor space wavefunctions
than those used in \wittwistors.  We take our particles to have
definite momenta $p_i^{a\dot a}=\lambda_i^a\tilde\lambda_i^{\dot
a}$ in Minkowski space.  The corresponding twistor space
wavefunction is\foot{The twistor space wavefunction is supposed to
be a $\bar\partial$-closed $(0,1)$-form with values in a line
bundle that depends on the helicity.  We have a
$\bar\partial$-closed
 $(0,1)$-form here because $\bar\delta(f)$, for any
holomorphic function $f$, is such a form. Since the line bundles
in question are naturally trivial when restricted to
$\lambda=\lambda_i$, we can (at the informal level of the present
discussion) write the wavefunctions without being very precise in
describing the line bundle.}
 \eqn\gurtu{\bar\delta(\langle\lambda,\lambda_i\rangle)
\exp(i[\mu,\tilde\lambda_i]).} The idea here is that this
wavefunction represents a particle of definite $\lambda$ because
the wavefunction has delta function support at
$\lambda=\lambda_i$, and it has definite $\tilde\lambda$ because
of the plane wave dependence on $\mu$.  Choosing twistor space
wavefunctions that represent momentum space eigenstates in
Minkowski space means that the twistor computation can be compared
directly to the standard momentum space scattering amplitudes,
without needing to perform an additional Fourier transform.   It
turns out that this also simplifies the computations.  (The same
simplification was achieved in \rsv\ by performing a Fourier
transform prior to evaluating the twistor scattering amplitude.)

The effective action for fields in twistor space is the integral
of a Chern-Simons $(0,3)$-form.  The kinetic operator for these
fields is the $\bar\partial$ operator.  The propagator is a
$(0,2)$-form on $\Bbb{CP}^3\times \Bbb{CP}^3$ that we write as
$G(\lambda',\mu'; \lambda,\mu)$, where $(\lambda,\mu)$ are
homogeneous coordinates for one point in $\Bbb{CP}^3$ and
$(\lambda',\mu')$ for the other. The part of $G$ that is a
$(0,1)$-form on each copy of $\Bbb{CP}^3$ is the propagator for
the physical fields, while as in quantization of real Chern-Simons
gauge theory \ref\axsing{S. Axelrod and I. M. Singer,
``Chern-Simons Perturbation Theory,'' hep-th/9110056, in {\it
Differential Geometric Methods In Theoretical Physics} (1991).},
the terms in $G$ that are $(0,2)$-forms on one $\Bbb{CP}^3$ and
$(0,0)$-forms on the other describe propagation of ghosts. We
write the equation that should be obeyed by $G$ in coordinates
with $\lambda^1=\lambda'{}^1=1$:\foot{The prefactor $1/2\pi$
depends on the proper normalization of the Chern-Simons
$(0,3)$-form action in twistor space.  We are making a guess based
on the analogous normalization for real Chern-Simons theory at
level one and will not try to prove that this is the correct
normalization of the propagator.} \eqn\onong{\bar\partial
G={1\over
2\pi}\bar\delta(\lambda'{}^2-\lambda^2)\bar\delta(\mu'^{\dot
1}-\mu^{\dot 1})\bar\delta(\mu'{}^{\dot 2}-\mu^{\dot 2}).} We can
therefore take the propagator to be \eqn\honong{G={1\over
(2\pi)^2}\bar\delta(\lambda'{}^2-\lambda^2)\bar\delta(\mu'{}^{\dot
1}-\mu^{\dot 1}){1\over \mu'{}^{\dot 2}-\mu^{\dot 2}}.} This
choice of $G$ amounts to a choice of gauge.

\ifig\tramppok{Twistor diagrams corresponding to MHV tree diagrams
that were considered in section 5.  There are two disconnected
instantons, labeled $C$ and $C'$, to which gluons are attached;
they are connected by a twistor space propagator.}
{\epsfbox{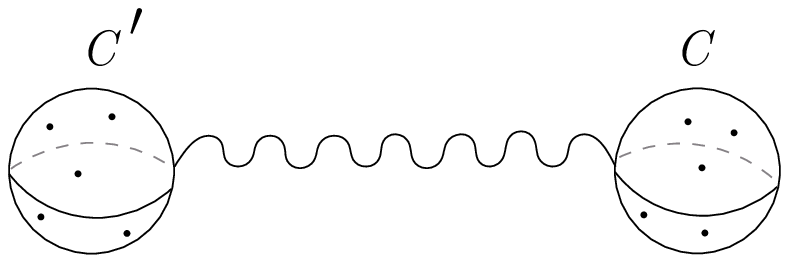}}
 Now consider the exchange of a twistor field between copies of
$\Bbb{CP}^1$ that represent instantons $C$ and $C'$ of degree one.
As in figure 10, the external gluons are attached to $C$ and $C'$.
$C$ is described by the equation \eqn\onsnng{\mu^{\dot a}=x^{a\dot
a}\lambda_a,} and $C'$ by the equation \eqn\bonsnng{\mu'^{\dot
a}=x'{}^{a\dot a}\lambda'_a.} We also set $y_{a\dot a}=x'_{a\dot
a}-x_{a\dot a}$.  $C'$ and $C$ will correspond respectively to the
vertices on the left and right of figure 7.

The exponential factors in \gurtu\ give an important dependence on
$x$ and $x'$.  Taking the product of the exponentials for all of
the external particles, we get \eqn\oyto{\prod_{i\in
L}\exp(ix'_{a\dot a}p_i^{a\dot a})\prod_{j\in R}
\exp(ix_{b\dot
b}p_j^{b\dot b}).} We can also write this expression as
\eqn\borto{ \exp(iy_{a\dot a}P^{a\dot a})\prod_i\exp(ix_{b\dot
b}p_i^{b\dot b}),} where as in section 5, $P=\sum_{i\in L}p_i$,
and in the second factor all external particles are included. The
integral over $x$ will give a delta function of energy-momentum
conservation; the $y$-dependent factor in \borto\ will also play
an important role.

We will take the measure for integrating over $x$ and $y$ to be
$d^4x^{a\dot a}\,\,d^4y^{b\dot b}$, where, for example,
$d^4y^{b\dot b}=dy^{1\dot 1}dy^{2\dot 2}dy^{2\dot 1}dy^{1\dot 2}$.

With our choice of gauge, in coordinates with
$\lambda^1=\lambda'{}^1=1$, the twistor propagator $G$ is
supported on pairs of points that obey $\lambda'{}^2=\lambda^2$.
We can more invariantly say simply that $\lambda'^a=\lambda^a$
(without specializing to coordinates with $\lambda^1=1$).  In
addition, as $\mu'^{\dot a}-\mu^{\dot a}=y^{a\dot a}\lambda_a$,
the condition that the propagator is exchanged between points with
$\mu'{}^{\dot 1}=\mu^{\dot 1}$ means that $y^{a\dot
1}\lambda_a=0$, or in other words that
\eqn\onnno{\lambda^a=y^{a\dot 1}} up to an irrelevant scaling. The
propagator contains a factor $1/(\mu'^{\dot 2}-\mu^{\dot
2})=1/y^{a\dot 2}\lambda_a=1/y^{a\dot 2}y_{a}{}^{\dot 1}$.  But
$y^{a\dot 2}y_{a}{}^{\dot 1}=\half(y^{a\dot 2}y_a{}^{\dot 1}-
y^{a\dot 1}y_a{}^{\dot 2})=-y^{a\dot a}y_{a\dot a}/2$.

So finally the integral representing the contribution $\tilde
I_\Gamma$ to the scattering amplitude from the instanton
configuration considered in figure 10 is \eqn\polko{\tilde
I_\Gamma=-{1\over 2\pi^2}\int {d^4y^{b\dot b}\over y^{a\dot
a}y_{a\dot a}}\exp(iy_{c\dot c}P^{c\dot c})g(\lambda;\lambda_i),}
where $\lambda^a=y^{a\dot 1}$, while $\lambda_i^a$ are the spinors
associated with external gluons.   The function
$g(\lambda;\lambda_i)$ arises from computing the correlation
function of gluon vertex operators on $C$ and $C'$ (and
integrating over fermionic moduli) as explained in section 4.7 of
\wittwistors. It is the same function that entered in section 5.
(The factor $\exp(iy_{a\dot a}P^{a\dot a})$ was absent in
analogous formulas in \wittwistors\ because different twistor
space wavefunctions were used.) Most of the ingredients
  in \polko\ are Lorentz-covariant; Lorentz covariance
is violated only because the function $g(\lambda;\lambda_i)$ is evaluated at
$\lambda^a=y^{a\dot 1}$, clearly a noncovariant condition.

We now have to decide how to interpret the integral in \polko. The
integrand is a holomorphic function of $y$ and the integral is a
complex contour integral of some sort.  We most definitely do not
know any systematic theory of how to pick the contours in
topological string theory in twistor space.  Here we will simply
describe a recipe for interpreting this integral that was found in
an attempt to match with our results about MHV tree diagrams.

We assume, first of all, that one of the $y$ integrals should be
performed via a contour integral around  the pole at
$y^2=0$, and thus gives $2\pi i$ times the residue of that pole.
The integral thus becomes an integral on the quadric $Q$ defined
by $y^2=0$. We write this schematically \eqn\hurfo{\tilde
I_\Gamma=-{i\over \pi}\int_Q {\rm Res}_{y^2=0}{d^4y^{b\dot b}\over
y_{c\dot c}y^{c\dot c}}\,\exp(iy_{a\dot a}P^{a\dot
a})g(\lambda,\lambda_i).} (We will compute this residue
momentarily.) Once this is done, our formula becomes
Lorentz-invariant. Indeed, at $y^2=0$, we can factor $y^{a\dot a}$
as $\lambda^a\tilde\lambda^{\dot a}$, where one way to determine
$\lambda$ is to say that up to an irrelevant scaling,
$\lambda^a=y^{a\dot 1}$.  In fact, the formula $y^{a\dot
a}=\lambda^a\tilde\lambda^{\dot a}$ implies $\lambda^a=y^{a\dot
1}/\tilde\lambda^{\dot 1}$.

We actually want to decompose $y^{a\dot a}$ a little differently.
We write \eqn\nonnns{y^{a\dot a}=t\lambda^a\tilde \lambda^{\dot
a},} where the $\lambda^a$ are homogeneous coordinates for one
copy of $\Bbb{CP}^1$, $\tilde\lambda^{\dot a}$ are homogeneous
coordinates for a second copy of $\Bbb{CP}^1$, and $t$ scales with
weight $-1$ under scaling of either $\lambda$ or $\tilde \lambda$.
The scaling of $t$ has been selected to ensure that $y$ is
invariant.  The measure on the quadric is determined by the
symmetries to be \eqn\nsnno{{\rm Res}_{y^2=0}{d^4y\over
y^2}=ft\,dt\,\langle\lambda,d\lambda\rangle
[\tilde\lambda,d\tilde\lambda ],} for some constant $f$ (which we
will soon find to equal $1/2$). The dependence on $\lambda$ and
$\tilde\lambda$ is determined from $SL(2)\times SL(2)$ invariance;
the power of $t$ can be fixed by requiring that the measure is
invariant under scaling of $\lambda$ or $\tilde\lambda$.

To compute $f$, we simply compare the two measures at a convenient
point $P$. The differential form $d^4y^{a\dot a}/y^{b\dot
b}y_{b\dot b}=dy^{1\dot 1}dy^{2\dot 2}dy^{2\dot 1}dy^{1\dot
2}/2(y^{1\dot 1}y^{2\dot 2}-y^{1\dot 2}y^{2\dot 1})$ has a pole at
$y^{2\dot 2}=y^{1\dot 2}y^{2\dot 1}/y^{1\dot 1}$ whose residue is
the volume form $\Phi = dy^{1\dot 1}dy^{2\dot 1}dy^{1\dot
2}/2y^{1\dot 1}$ on $Q$. The point $P$ at which the only nonzero
component of $y$ is $y^{1\dot 1}=1$ corresponds in the other
variables to $t=1$, $\lambda^a=(1,0)$, $\tilde\lambda^{\dot
a}=(1,0)$. Expanding around this point, we take $t=1+\epsilon$,
$\lambda^a=(1,\beta)$, $\tilde\lambda^{\dot a}=(1,\gamma)$, whence
to first order $y^{1\dot 1}=1+\epsilon,$ $y^{2\dot 1}=\beta$,
$y^{1\dot 2}=\gamma$.  So at $P$,
$\Phi=d\epsilon\,d\beta\,d\gamma/2$.  On the other hand,
$dt=d\epsilon$, $\vev{\lambda,d\lambda}=d\beta$,  and
$d\tilde\lambda=d\gamma$. So
$t\,dt\vev{\lambda,d\lambda}[\tilde\lambda,d\tilde\lambda]
=d\epsilon\,d\beta\,d\gamma$. Comparing these formulas, we find
that $f=1/2$.

 Our integral therefore becomes
\eqn\invo{\tilde I_\Gamma=-{i\over 2\pi}\int t\,dt \langle
\lambda,d\lambda\rangle[\tilde\lambda,d\tilde\lambda]\exp(it\lambda_a\tilde
\lambda_{\dot a}P^{a\dot a}) ~g(\lambda;\lambda_i).} Again, the
proper interpretation of this integral is unclear.  Trying to get
an answer that makes some sense, we interpret the $t$ integral as
an integral from 0 to $\infty$, using \eqn\unon{\int_0^\infty t\,
dt\,\exp(i\gamma t)=-{1\over \gamma^2}.} So \eqn\binvo{\tilde
I_\Gamma ={i\over 2\pi}\int \langle\lambda,d\lambda\rangle
[\tilde\lambda,d\tilde\lambda]{1\over (P_{a\dot
a}\lambda^a\tilde\lambda^{\dot a})^2} g(\lambda;\lambda_i).} Thus
we have motivated the integral that we took as our starting point
in section 5.\foot{The fact that the coefficient of the integral
has come out correctly is somewhat fortuitous, as we have not been
precise enough with our twistor space calculation to be certain of
an absolute multiplicative factor.}

Integrating over $t$ from 0 to $\infty$ seems rather unpalatable
in the context of complex contour integrals, since contours are
normally closed cycles or cycles that run off to infinity (rather
than terminating at $t=0$).  The procedure that we have followed
seems somewhat more plausible in conjunction with the choice we
made in section 5 of setting $\tilde\lambda=\bar\lambda$.  With
$y^{a\dot a}=t\lambda^a\tilde\lambda^{\dot a}$, the combined
operation of setting $\tilde\lambda=\bar\lambda$ and taking $t$ to
be real and positive amounts to integrating over the future light
cone in real Minkowski spacetime; this seems like a more or less
respectable integration cycle, albeit singular at the origin and
noncompact.

The proof of $SL(2,\Bbb{C})$ invariance at the end of section 5
shows that integration over the past light cone would give the
same result.  In fact, unlike the real light cone, which has a
future and a past, the complexified light cone is connected.  We
have shown that our amplitudes, after summing over graphs, are
invariant under the complexified Lorentz group.  This group can be
used to rotate the future real light cone to the past real light
cone.

What shall we make of our result?  If our procedure for calculating the
integral is correct, then the tree level Yang-Mills scattering
amplitudes appear to come from totally disconnected instanton
configurations in twistor space.  On the other hand, there appears
to be convincing evidence \rsv\ that they can be computed from
connected instantons alone.  Are there really two distinct ways to
compute the same amplitudes from twistor space? For more general
amplitudes, are there more than two ways, allowing for instantons
of higher degree that are neither connected nor completely
disconnected?  Or is there a fault in the way the integrals have
been evaluated? Certainly, we cannot claim a firm justification for the
way that we have evaluated the integral.  So in
our computation, there are ample possibilities to suppose that a
more complete and rigorous evaluation of the scattering amplitude
might require including additional contributions.

\vskip 2 cm

Work of F. Cachazo was supported in part by the Martin A. and
Helen Chooljian Membership at the Institute for Advanced Study and
by DOE grant DE-FG02-90ER40542; that of P. Svrcek by NSF grants
PHY-9802484 and PHY-0243680; and that of E. Witten by NSF grant
PHY-0070928. Opinions and conclusions expressed here are those of
the authors and do not necessarily reflect the views of funding
agencies.
 \listrefs
\end